\documentclass[twoside,twocolumn,9pt]{article}
\usepackage{extsizes}
\usepackage[none]{hyphenat}
\usepackage[super,sort&compress,comma]{natbib} 
\usepackage[version=3]{mhchem}
\usepackage[left=1.5cm, right=1.5cm, top=1.785cm, bottom=2.0cm]{geometry}
\usepackage{balance}
\usepackage{stfloats}
\usepackage{mathptmx}
\usepackage{sectsty}
\usepackage{graphicx} 
\usepackage{lastpage}
\usepackage[format=plain,justification=justified,singlelinecheck=false,font={stretch=1.125,small,sf},labelfont=bf,labelsep=space]{caption}
\usepackage{float}
\usepackage{fancyhdr}
\usepackage{fnpos}
\usepackage[english]{babel}
\addto{\captionsenglish}{%
  
}
\usepackage{array}
\usepackage{droidsans}
\usepackage{charter}
\usepackage{multirow}
\usepackage[T1]{fontenc}
\usepackage[usenames,dvipsnames]{xcolor}
\usepackage{setspace}
\usepackage[compact]{titlesec}
\usepackage{hyperref}
\usepackage{epstopdf}%This line makes .eps figures into .pdf - please comment out if not required.

\definecolor{cream}{RGB}{222,217,201}

\begin{document}

\pagestyle{fancy}
\thispagestyle{plain}
\fancypagestyle{plain}{
%%%HEADER%%%
\renewcommand{\headrulewidth}{0pt}
}
%%%END OF HEADER%%%

%%%PAGE SETUP - Please do not change any commands within this section%%%
\makeFNbottom
\makeatletter
\renewcommand\LARGE{\@setfontsize\LARGE{15pt}{17}}
\renewcommand\Large{\@setfontsize\Large{12pt}{14}}
\renewcommand\large{\@setfontsize\large{10pt}{12}}
\renewcommand\footnotesize{\@setfontsize\footnotesize{7pt}{10}}
\makeatother

\renewcommand{\thefootnote}{\fnsymbol{footnote}}
\renewcommand\footnoterule{\vspace*{1pt}% 
\color{cream}\hrule width 3.5in height 0.4pt \color{black}\vspace*{5pt}} 
\setcounter{secnumdepth}{5}

\makeatletter 
\renewcommand\@biblabel[1]{#1}            
\renewcommand\@makefntext[1]% 
{\noindent\makebox[0pt][r]{\@thefnmark\,}#1}
\makeatother 
\renewcommand{\figurename}{\small{Fig.}~}
\sectionfont{\sffamily\Large}
\subsectionfont{\normalsize}
\subsubsectionfont{\bf}
\setstretch{1.125} %In particular, please do not alter this line.
\setlength{\skip\footins}{0.8cm}
\setlength{\footnotesep}{0.25cm}
\setlength{\jot}{10pt}
\titlespacing*{\section}{0pt}{4pt}{4pt}
\titlespacing*{\subsection}{0pt}{15pt}{1pt}
%%%END OF PAGE SETUP%%%

%%%FOOTER%%%
\fancyfoot{}
\fancyfoot[LO,RE]{\vspace{-7.1pt}\includegraphics[height=9pt]{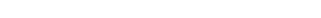}}
\fancyfoot[CO]{\vspace{-7.1pt}\hspace{13.2cm}\includegraphics{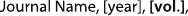}}
\fancyfoot[CE]{\vspace{-7.2pt}\hspace{-14.2cm}\includegraphics{head_foot/RF}}
\fancyfoot[RO]{\footnotesize{\sffamily{1--\pageref{LastPage} ~\textbar  \hspace{2pt}\thepage}}}
\fancyfoot[LE]{\footnotesize{\sffamily{\thepage~\textbar\hspace{3.45cm} 1--\pageref{LastPage}}}}
\fancyhead{}
\renewcommand{\headrulewidth}{0pt} 
\renewcommand{\footrulewidth}{0pt}
\setlength{\arrayrulewidth}{1pt}
\setlength{\columnsep}{6.5mm}
\setlength\bibsep{1pt}
%%%END OF FOOTER%%%

%%%FIGURE SETUP - please do not change any commands within this section%%%
\makeatletter 
\newlength{\figrulesep} 
\setlength{\figrulesep}{0.5\textfloatsep} 

\newcommand{\topfigrule}{\vspace*{-1pt}% 
\noindent{\color{cream}\rule[-\figrulesep]{\columnwidth}{1.5pt}} }

\newcommand{\botfigrule}{\vspace*{-2pt}% 
\noindent{\color{cream}\rule[\figrulesep]{\columnwidth}{1.5pt}} }

\newcommand{\dblfigrule}{\vspace*{-1pt}% 
\noindent{\color{cream}\rule[-\figrulesep]{\textwidth}{1.5pt}} }

\makeatother
%%%END OF FIGURE SETUP%%%

%%%TITLE, AUTHORS AND ABSTRACT%%%
\twocolumn[
  \begin{@twocolumnfalse}
%{\includegraphics[height=30pt]{head_foot/SM}\hfill\raisebox{0pt}[0pt][0pt]{\includegraphics[height=55pt]{head_foot/RSC_LOGO_CMYK}}\\[1ex]
%\includegraphics[width=18.5cm]{head_foot/header_bar}}\par
\vspace{1em}
\sffamily
\begin{tabular}{m{0\textwidth} p{0.95\linewidth} }

\multirow{2}{*} & \noindent\LARGE{\textbf{Multi-field induced collective behavior and tunable assembly of capsule-shaped Hematite particles }} \\%Article title goes here instead of the text "This is the title"
\vspace{0.3cm} & \vspace{0.3cm} \\

 & \noindent\large{Jehan Daftari,\textit{$^{ab}$} Indira Barros,\textit{$^{a}$} and Indrani Chakraborty$^{\ast}$\textit{$^{a}$}} \\%Author names go here instead of "Full name", etc.

& \noindent\normalsize{We investigate the multi-field directed assembly of dielectric, canted antiferromagnetic hematite (\(\alpha-Fe_2O_3\)) capsules, exploiting their unique dual responsiveness to electric and magnetic fields. Under AC electric fields with a frequency range of 5-50 kHz, electric dipolar torques induce an alignment of the capsule long axes from a parallel to a perpendicular orientation with respect to the field direction. At a higher frequency of 500 kHz, the capsules aggregate into bundles due to attractive magnetic interactions, weakening electric dipolar repulsion and the appearance of electrothermal flow. Under DC magnetic fields, the capsules produce side-by-side ribbons indicating a magnetic dipole moment vector situated along the capsule waist. When a combined electric and magnetic field is applied in parallel, the capsules show a cooperation between electric and magnetic torques at 5 kHz and a competition between them at 50 kHz. On perpendicular application of the electric and magnetic fields, independent contributions of the electric/magnetic torques determine the final assembly. A combination of Clausius-Mossotti factor calculations and finite element modelling is used to explain our observations. By tuning the pH of the medium and the applied field frequency, explicit demonstration of the cargo trapping potential of the hematite capsules are shown, thereby highlighting their potential in fabricating stimuli-sensitive smart materials and cargo trapping and delivery systems.

}

\end{tabular}

 \end{@twocolumnfalse} \vspace{0.6cm}

  ]
%%%END OF TITLE, AUTHORS AND ABSTRACT%%%

%%%FONT SETUP - please do not change any commands within this section
\renewcommand*\rmdefault{bch}\normalfont\upshape
\rmfamily
\vspace{-1cm}

%%%FOOTNOTES%%%

\footnotetext{\textit{$^{a}$~Department of Physics, Birla Institute of Technology and Science, Pilani, K K Birla Goa Campus, Zuarinagar, Sancoale, Goa 403726, India.}}
\footnotetext{\textit{$^{b}$~Department of Electrical \& Electronics Engineering, Birla Institute of Technology and Science, Pilani, K K Birla Goa Campus, Zuarinagar, Sancoale, Goa 403726, India.}}
\footnotetext{\textit{$^*$~E-mail: indranic@goa.bits-pilani.ac.in}}
%Please use \dag to cite the ESI in the main text of the article.
%If you article does not have ESI please remove the the \dag symbol from the title and the footnotetext below.
%\footnotetext{\dag~Supplementary Information available: [details of any supplementary information available should be included here]. See DOI: 10.1039/cXsm00000x/}
%additional addresses can be cited as above using the lower-case letters, c, d, e... If all authors are from the same address, no letter is required

%\footnotetext{\dag~Electronic supplementary information (ESI) available. }

%\ddag for both dagguers amd reintroduce the part about the SI
%%%END OF FOOTNOTES%%%

%%%MAIN TEXT%%%%
\section{Introduction}
Programmable self-assembly of reconfigurable structures using colloidal `Lego' blocks has become an area of intense research over the past decade owing to its intense application potential in producing stimuli-sensitive smart materials, switchable photonic crystals, microswimmers for targeted drug delivery, mechanical metamaterials, electronics beyond conventional lithography and advanced environmental sensors.\cite{barros2026field, AlHarraq2022, Wei2026, landers2025clinically, Zhu2025, Manoharan2015, Glotzer2007, Sacanna2011} While a wide range of techniques, including DNA-mediated assembly\cite{chakraborty2022self,shelke2025self, van2024dna, rogers2016using}, patchy particle-based assembly\cite{Sacanna2011, palacci2013living, barros2026field}, and hydrophobic-hydrophilic interaction-based assembly\cite{rinaldin2019colloid}, have been implemented to introduce tunability and structural diversity in several colloidal systems, the complexity of artificial structures achievable with engineered building blocks remains far below that observed in natural systems like a snowflake or a virus capsid\cite{aubret2018targeted, Wei2026, zhao2026effect}. One of the simplest, most cost-effective, and widely employed approaches for creating adaptable and complex assemblies is field-directed assembly using electric, magnetic, or optical fields\cite{Barros2025SoftMatter,barros2026cooperation, barros2026field, Ketzetzi2025,Ma2013, Ma2015, aubret2018targeted, palacci2013living, Gong2017, Zhu2025, Haque2025}. Combining multi-field excitation with shape-anisotropic colloidal building blocks can lead us to achieving smart, reconfigurable structures with unprecedented structural reconfigurability.\cite{barros2026cooperation,barros2026field, Ma2013, Ma2015, Gong2017, Zhu2025, Haque2025, edwards2014controlling, Crassous2014} 

Among field-based colloidal assembly techniques, electric-field-based assembly of isotropic and anisotropic colloidal particles has been employed extensively owing to its ease of implementation and low cost\cite{Barros2025SoftMatter, barros2026cooperation, Ma2013, Ma2015, Gong2017, Zhu2025, Haque2025, zhao2026effect, Ketzetzi2025, kwaadgras2014orientation, edwards2014controlling, Crassous2014, emerse2024morphologies}. Even then, unexpected structural configurations emerge frequently even from the simplest systems-for example, the formation of oligomers and non-close-packed crystalline lattices from polystyrene spheres\cite{Ma2013, yang2019impact}. This is because electric field-induced assembly is governed by a subtle interplay of several mechanisms including dipole--dipole interactions, dielectrophoretic (DEP) forces, electrohydrodynamic (EHD) flows, induced-charge electro-osmosis (ICEO) and concentration-polarization electro-osmosis (CPEO), whose relative importance depends sensitively on frequency, voltage, particle size, electrolyte composition, zeta potential and other particle and electrode surface properties \cite{Barros2025SoftMatter, barros2026cooperation, yang2019impact}. Magnetic fields, by comparison, avoid much of this complexity because the induced interactions are purely dipolar in origin, although this simplicity generally comes at the expense of a comparatively narrower range of accessible structures \cite{Bharti2016, AlHarraq2022}. At the same time, magnetic field-based assembly offers greater practical applicability owing to its penetrability through biological tissues and its lower sensitivity to the environmental conditions of the concerned system \cite{landers2025clinically}. A combination of electric- and magnetic-field-based excitation would therefore open up a whole new regime of diverse, reconfigurable structures with a high degree of tunability and even formation of structural hierarchy. Using spherical particles, Bharti \textit{et al.} demonstrated that concurrent electric and magnetic fields produce bidirectional chains, networks, and discrete crystalline structures that neither field alone can produce \cite{Bharti2016}. More recently, Haque \textit{et al.} employed orthogonal electric and magnetic fields to independently control attractive and repulsive dipolar interactions, producing dense, well-aligned linear chains that subsequently organized into hierarchical superstructures at higher particle concentrations \cite{Haque2025}. Barros \textit{et al.} further demonstrated that the interplay between electric and magnetic fields can be programmed by controlling the electric field frequency and field orientation\cite{barros2026cooperation}. The introduction of anisotropic particles into a multi-field-based excitation protocol introduces an even greater degree of complexity \cite{barros2026field}. In an applied electric field, the induced dipole moment of a non-spherical particle depends on its instantaneous orientation relative to the field, thereby intrinsically coupling rotational and translational degrees of freedom. Moreover, unlike in a spherical colloidal system, the surrounding electrohydrodynamic flow field is generally non-axisymmetric, giving rise to a significantly richer and more complex landscape of interparticle interactions. For example, Wei \textit{et al. }recently demonstrated that chains of flexible anisotropic microstructures actuated by AC electric fields exhibit a diverse range of emergent behaviors, including undulatory swimming, autonomous rotation, and collision avoidance, arising solely from the coupling between particle shape, flexibility, and the applied field, without requiring sensing or feedback control \cite{Wei2026}. Anisotropic magnetic colloids, on the other hand, lead to the formation of a diverse range of structures, from arrays of magnetic cubes to columnar chains, in which the easy axis of magnetization becomes a decisive factor in the structural arrangement.\cite{kamal2020path, rossi2018self, sudhakaran2020orientation, sacanna2013engineering, sacanna2012magnetic, tierno2014recent} \\

Capsule-shaped hematite colloids offer a unique system that is responsive to both electric and magnetic fields, owing to their dielectric and canted antiferromagnetic nature. The magnetic easy axis of hematite ellipsoids has been observed to lie within the basal plane and is therefore perpendicular to the particle's geometric long axis\cite{emerse2024morphologies}. Consequently, under an applied magnetic field, these prolate particles align with their short axis parallel to the field. This coupling between particle shape and magnetic anisotropy gives rise to liquid-crystalline phases, including a previously unreported smectic phase, that cannot be predicted from particle geometry alone \cite{Martchenko2016, Liao2022}. Under an electric field, hematite and other ellipsoidal colloids have been reported to assemble into well-defined tubular and sheet-like superstructures whose morphology depends sensitively on particle aspect ratio \cite{Crassous2014}. To the best of our knowledge, no report exists on the multi-field tunable assembly of hematite ellipsoids/capsules, even though that is expected to produce a richer phase space of structural configurations, nor are the relative roles of the magnetic and electric forces/torques explored in the same physical system having a specific shape. In this work, we fabricate sphero-cylinder-shaped hematite capsules and explore their response and assembly under single-field and multi-field excitation. By systematically varying the frequency, field strength and field direction, we identify distinct assembly regimes governed by the nuanced interactions between electric and magnetic dipolar forces, dipolar torques and electrohydrodynamic flows. Using a set of sequential experiments, finite element modelling and Clausius-Mossotti factor calculations, we bring out the competition, cooperation and independent contributions of the electric and magnetic forces/torques in determining the final structural configuration in a multi-field responsive system.   

\begin{figure*}[!h]
\centering
  \includegraphics[width=16.0 cm, keepaspectratio]{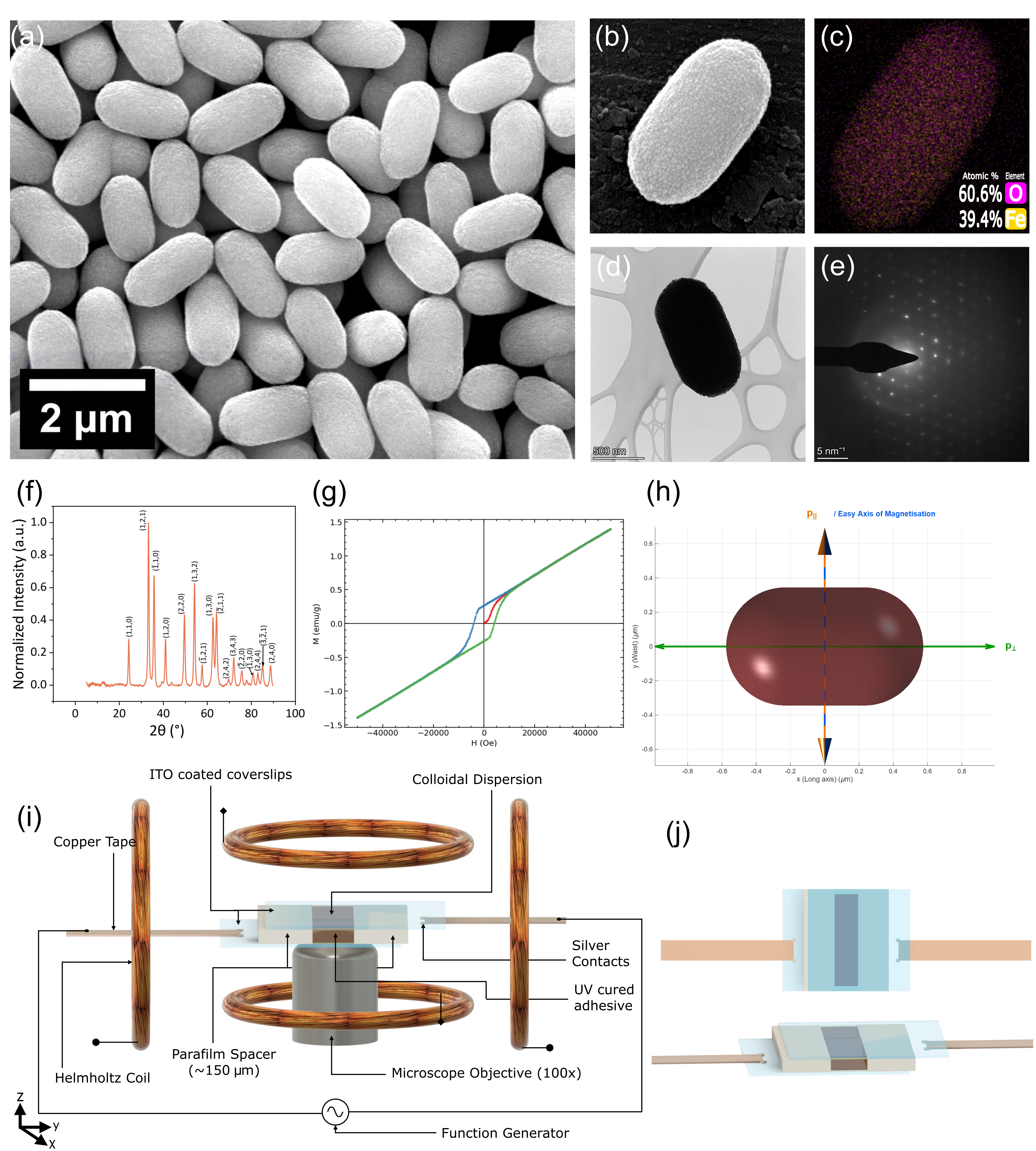}
  \caption{(a) Scanning electron microscopy (SEM) image of synthesized hematite capsules ($\alpha-Fe_2O_3$). (Scale bar, 2 $\mu m$.) (b) SEM image of a single hematite capsule. (c) Energy-dispersive X-ray spectroscopy (EDS) shows Fe and O contents in terms of atom \%: Fe 39.4\% and O 60.6\%) (d) Transmission electron microscope (TEM) image of one capsule shows multiple crystallites along the capsule edge. (e) Selected area diffraction pattern for a crystal on one of the corners of the capsule shows the presence of spots as well as diffused rings indicating polycrystallinity. (f) X-ray diffraction pattern for the hematite capsules indicates the trigonal crystal structure of $\alpha$-Fe$_2$O$_3$ (space group $R\bar{3}c$) with the two most intense peaks representing the (121) and ($\bar{1}10$) set of planes. (g) Magnetisation vs applied magnetic field (M-H) curve for the hematite capsules at 300 K. The residual magnetization of 0.26 emu/g indicates the canted antiferromagnetic nature of hematite. (h) Schematic diagram showing the alignment of the magnetic and electric dipole moment vectors in a hematite capsule. (i) Schematic diagram of the experimental set-up showing the Helmholtz coil placements for applying the magnetic field in the z or x-y directions. (j) Schematic of the sample chamber showing the skewed placement of the top electrodes and the capillary space created for the colloidal sample.}
  \label{characterizations}
\end{figure*}

\section{Experimental section}
\subsection{Synthesis and characterization} 
Capsule-shaped hematite ($\alpha$-Fe$_2$O$_3$)  particles were prepared by a modified version of the sol-gel method as outlined by Sugimoto \textit{et al.}\cite{sugimoto1998shape} 90 mL of 6 M sodium hydroxide (NaOH) solution was added to 100 mL of 2 M freshly obtained ferric chloride (FeCl$_3$.6H$_2$O) in a borosilicate glass bottle. This was followed by the addition of 10 mL of 0.3 M Sodium dodecyl sulphate (SDS) solution. The mixture turned reddish-brown and was then shaken vigorously to homogenize the gel. The bottle was tightly closed with a plastic cap, and the gel was left undisturbed at 100$^\circ$C for 8 days in a pre-heated oven to allow the reaction to complete. After 8 days, the bottle was taken out, cooled to room temperature and the particles were washed via a series of sequential sedimentation and re-suspension cycles in deionized water. The zeta potential of the particles measured in deionized water at 25$^\circ$C was -27 mV, indicating fair stability. The morphology of the hematite capsules was observed under a field emission scanning electron microscope (Quanta FEG 250). The FE-SEM micrographs revealed monodisperse, micrometer-sized hematite capsules (Fig. \ref{characterizations}a and b) with a mean major axis of 1.08 $\pm$ 0.20 $\mu$m and a mean minor axis of 0.74  $\pm$ 0.21 $\mu$m. Energy-dispersive X-ray microanalysis (EDS) showed a uniform distribution of Fe and oxygen with atomic concentrations of 39.4\% Fe and 60.6\% Oxygen. (Fig. \ref{characterizations}c). The calculated stoichiometric ratio of Fe:O (1:1.5) is consistent with the successful synthesis of the hematite phase.  Transmission electron microscope (TEM) image of a single hematite capsule (Fig. \ref{characterizations}d) showed the arrangement of crystallites at the edge of the capsule, but the particle was otherwise opaque to the electron beam. A selected area diffraction pattern (SAED) (Fig. \ref{characterizations}e) taken from the edge of the capsule showed a superposed pattern consisting of diffraction spots and rings, indicating the presence of multiple crystallites, even within a single particle. For the same sample, powder X-ray diffraction (XRD) plot (Fig. \ref{characterizations}f) indicated the trigonal crystal structure of $\alpha$-Fe$_2$O$_3$ (space group $R\bar{3}c$) with the two most intense peaks representing the (121) and ($\bar{1}10$) set of planes. The magnetization vs applied field (M-H) measurements at room temperature (300K/25$^\circ $C) were done using a vibration sample magnetometer (VSM). The appearance of a hysteresis loop in the M-H plot confirms the expected canted antiferromagnetic nature of hematite previously observed at room temperature.\cite{cantedPhysRev.116.1447} Within an applied magnetic field range of -5T to +5T, the hematite capsules did not show any saturation (Fig. \ref{characterizations}g). The residual magnetism ($M_r$) was measured to be 0.26 emu/g while the coercivity ($H_c$) was measured to be 400 mT. For ellipsoidal hematite particles, the magnetic dipole moment vector has been shown to lie along the waist of the ellipsoid, parallel to the short axis,\cite{martinez-10.1039/c5sm02823j} while the electric dipole moment vector has components parallel and perpendicular to the long and short axes respectively (Fig. \ref{characterizations}h))

\subsection{Sample chamber preparation} \label{setup-desc}
The experimental chamber was fabricated as a parallel-plate capillary cell utilizing two Indium Tin Oxide (ITO) coated cover-slips as the electrodes. To facilitate accessible electrical contact points, the electrodes were assembled in an offset configuration (see Fig. \ref{characterizations}i and j). A well-defined inter-electrode gap of 150 $\mu m$ was established using a Parafilm spacer, which served two purposes: maintaining the capillary's depth and preventing the electrodes from contacting and shorting. Electrical connectivity was achieved via copper tape leads, with interfacial resistance minimized by applying conductive silver paste. The colloidal dispersion was introduced into the cell by carefully pipetting the sample into the capillary. To prevent solvent evaporation, leakage, and hydrodynamic drift, all open ends of the setup were sealed using a UV-curable adhesive (Norland Optical Adhesive No. 81).

\subsection{Field-based experiments and imaging} 
The colloidal system was subjected to sinusoidal AC electric fields ($E$) using a METRAVI DDS-1010 function generator, and a Helmholtz coil assembly was used to create a uniform magnetic field ($B$) over the sample. The magnetic field was configured either parallel ($B \parallel E$) or perpendicular ($B \perp E$) to the electric field vector (Fig. \ref{characterizations}i)). Between consecutive observations, a time interval was kept to ensure the re-dispersion of particles by Brownian diffusion and the dissipation of any charge buildup, thereby restoring the system to a randomized state. The field-driven assembly of the hematite capsules was observed under an inverted microscope in bright-field mode, using a 100$\times$ objective and a high-speed camera.

\section{Results}

To systematically investigate the field-directed assembly of the hematite capsules, we first examine their response to AC electric fields in the absence of an external magnetic field, followed by their assembly under purely magnetic fields, and finally explore their behavior under combined electric and magnetic field excitations.

\subsection{Assembly of hematite capsules under AC Electric Fields}

\begin{figure*}[!h]
\centering
  \includegraphics[width=18.0 cm, keepaspectratio]{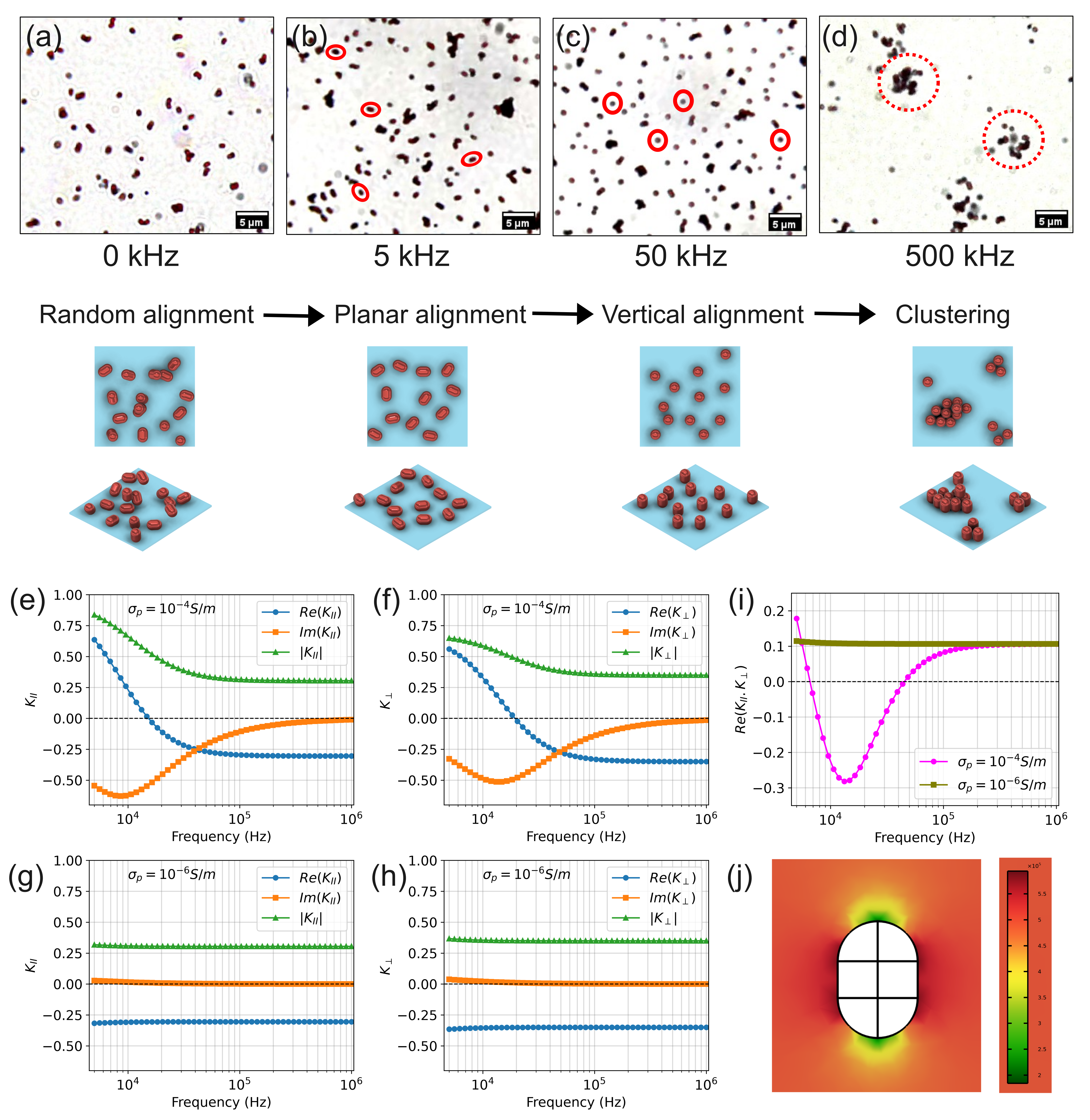}
  \caption{Assembly of hematite capsules under an AC electric field. (a) Random distribution of the capsules under no-field condition. (b) The capsules go into a planar arrangement with their long axes parallel to the observation plane at 5 kHz. A few capsules are marked in red ellipses. (c) The capsules arrange into a vertically aligned configuration (along the field) at a frequency of 50 kHz. A few vertically aligned capsules are marked in red circles. (d) At 500 kHz, the capsules form clustered bundles (marked in red dashed circles). The schematic diagrams given below indicate the arrangement of the capsules as observed from the top (top row) and from the side (bottom row). Calculation of the Clausius-Mossotti factors over a frequency range of 5 kHz-1 MHz, and for a hematite conductivity of $\sigma_p = 10^{-4} S/m$ along (e) the long axes ($K_{II}$) and (f) the short axes of the capsules ($K_{\perp}$). The same calculations for hematite conductivity $\sigma_p = 10^{-4} S/m$ are shown in (g) and (h). (i) A plot of the factor $Re(K_{II}.K_{\perp})$ as a function of the frequency shows that this factor and, hence, the electric dipolar torque become negative between 6 kHz and 50 kHz. (j) Finite element modelling of a single capsule with its long axis oriented along the electric field (parallel to the z-axis) shows that the highest magnitude of the field is in the equatorial regions while the lowest magnitude is at the poles.}
  \label{assembly_e}
\end{figure*}

The response of the hematite capsules to an applied AC electric field was studied over a frequency range of 1-500 kHz.  When no electric field is applied, the capsules are distributed in random orientations over the viewing plane (Fig. \ref{assembly_e}a). At a frequency of 5 kHz, the capsules arrange into a planar configuration parallel to the viewing plane (Fig. \ref{assembly_e}b). Beyond 5 kHz, the capsules begin to stand up, that is, orient their long axis along the electric field in the z direction, and by 50 kHz, all the capsules are oriented in the vertical direction (Fig. \ref{assembly_e}c). Finally, by 500 kHz, the capsules arrange into clustered configurations, as shown in Fig. \ref{assembly_e}d (we show these frequencies because the structures are fully formed and stable at these frequencies; the transitions themselves occur gradually at lower frequencies). To understand this behavior, we have to consider the relative roles of magnetic and electric dipolar interactions, as well as electrohydrodynamic (EHD) forces. Additionally, since the hematite capsule has an anisotropic shape, not only forces but also torques will be generated by dipolar interactions on these particles, thereby arranging them into specific configurations. If we represent the applied AC electric field with frequency $\omega$ as the real part of a complex phasor $\textbf{E}(t) = E_0 e^{i\omega t}\hat{e_z}$, then the time-averaged (electric) dipole-dipole interaction force on a spherical particle at position \textbf{r} exerted by another spherical particle residing at the coordinate origin can be written as: \cite{Gong2017, barros2026cooperation, Barros2025SoftMatter}
 \begin{equation}
     \textbf{F}_{E}(\textbf{r})=\frac{3}{4}\pi\epsilon _m a^2K_0^2E_{rms}^2 \left(\frac{2a}{r}\right)^4 \left[(1-3\cos^2{\theta})\hat{r}+\sin{2\theta}\hat{\theta}\right]  
\label{fdip}    
\end{equation}

Here, $K_0$ is the magnitude of $K$, which is the frequency-dependent complex Clausius–Mossotti (CM) factor. The CM factor is expressed as $K=\frac{\tilde{\epsilon}_p-\tilde{\epsilon}_m}{\tilde{\epsilon}_p+2\tilde{\epsilon}_m}$, where $\tilde{\epsilon}_p$ and $\tilde{\epsilon}_m$ are the complex permittivities of the particle and the suspending medium, respectively. Here $\epsilon_m$ is the medium permittivity, $a$ is the particle radius, $\textbf{r}$ represents the centre-to-centre separation between particles, $E_{rms}$ is the root-mean-square (rms) value of the applied electric field, and $\theta$ is the angle between the interparticle axis and the applied field. For $\theta = 0^\circ$, dipolar interactions between the particles are attractive, whereas for $\theta = 90^\circ$, the interactions are repulsive. On the other hand, the expression for the EHD force field surrounding a spherical colloidal source particle situated close to an electrode is given as:\cite{Ma2015, ristenpart2004assembly, barros2026cooperation}

\begin{equation}
F_{EHD} = \beta \left(\frac{9}{2}\pi \varepsilon_m a\right)\,\kappa h E_{\mathrm{rms}}^2
\frac{
Re(K) + \bar{\omega} Im(K)
}{
1 + \bar{\omega}^2
}
\frac{\left(\frac{r}{a}\right)}{\left[1 + \left(\frac{r}{a}\right)^2\right]^{5/2}}
\end{equation}

where $\bar\omega = \frac{\omega h}{2 \kappa D}$ is the scaled frequency, $\kappa^{-1}$ is the Debye screening length, D is the ion diffusivity, $h$ is the separation between the two electrodes, and $\beta=0.04$ is a scaling factor. $Re(K)$ and $Im(K)$ represent the real and imaginary parts of the CM factor, respectively.\cite{Ma2015} The tangential flow velocity, and hence the EHD force, is dependent on the frequency ($\sim \omega ^{-1}$) of the applied $E$ field and at lower frequencies, therefore ($\sim$ 5-20 kHz), the EHD force is dominant over the dipolar force. 
For the case of the capsule shaped particles, shape anisotropy will strongly influence the assembly behavior under electric and magnetic fields. The important point that has to be considered here is the fact that each capsule has a residual magnetization along its short axis and also carries an electric dipole moment which can be split into two components parallel and perpendicular ($p_{II}$ and $p_{\perp}$) to the long axis of the capsule\cite{jones1995electromechanics}. Assuming the capsules to be approximate prolate ellipsoids with similar major and minor axes, we can write:
\begin{equation}
 p_{II}=4\pi\left(\frac{L+2R}{2}\right)R^2\varepsilon_m K_{II}E_{II},
 \qquad
p_{\perp}=4\pi\left(\frac{L+2R}{2}\right)R^2\varepsilon_m K_{\perp}E_{\perp} 
\end{equation}
where $K_{II}$ and $K_{\perp}$ are the two components of the CM factor parallel and perpendicular to the long axis, respectively. Here the CM factor for an ellipsoid along a principal axis $n$ is given as:\cite{jones1995electromechanics}

\begin{equation}
    K_{n}(\omega) = \frac{\tilde{\epsilon}_p-\tilde{\epsilon}_m}{3(\tilde{\epsilon}_m+A_n (\tilde{\epsilon}_p-\tilde{\epsilon}_m))}
\end{equation}

where $A_n$ is the depolarization factor for the axis $n$ given as:
\begin{equation}
    A_n = \frac{a_1a_2a_3}{2}\int_{0}^{\infty} \frac{ds}{(s+a_n^2)\sqrt{(s+a_1^2)(s+a_2^2)(s+a_3^2)}}
\end{equation}

$s$ being an arbitrary distance for integration. For the hematite capsules, the sem-major axis $a_1=(L+2R)/2 = 0.54 \mu$m while the semi-minor axes $a_2=a_3=0.37 \mu$m. From this, the depolarization factor can be calculated to be $A_1=0.24$ and $A_2=A_3=0.38$. From this, the values of $K_{II}$ and $K_{\perp}$ can be calculated as shown in Fig. \ref{assembly_e}e-h. The calculation of the Clausius-Mossotti factors is sensitive to the accurate estimation of the conductivity of hematite. In literature, hematite conductivity is reported to vary from $10^{-4} $ to $ 10^{-9}$ S/m depending on several factors such as crystallinity, defect concentration, nature of the solvent, and particle size/shape. It is to be noted that to accurately measure the hematite conductivity, a measurement of the conductivity of an individual capsule in air/vacuum is not enough, as the presence of the electric double layer and the ion concentration of the surrounding medium will influence the conductivity once immersed inside a liquid. Hence, we calculated the CM factors using two limiting values of hematite conductivity above ($\sigma_p=10^{-4}$ S/m) and below ($\sigma_p=10^{-6}$ S/m) the conductivity of the medium (water) which we take as $\sigma_m=10^{-5}$ S/m. The dielectric constants of hematite and water are taken as $\varepsilon_p=20$ and $\varepsilon_m=80$, respectively. We plotted $K_{II}$ and $K_{\perp}$ for both $\sigma_p=10^{-4}$ S/m and $\sigma_p=10^{-6}$ S/m as shown in Fig. \ref{assembly_e}e-f and \ref{assembly_e}g-h respectively. While at $\sigma_p=10^{-6}$ S/m, the real part of both the  $K_{II}$ and $K_{\perp}$ were negative indicating negative dielectrophoresis, for $\sigma_p=10^{-4}$ S/m, both $K_{II}$ and $K_{\perp}$ showed a transition from positive to negative between 10 and 20 kHz. The absolute value of the CM factor showed a decay and subsequent saturation beyond 100 kHz. While a positive dielectrophoresis indicates that the capsules will tend to move towards high field regions, a negative dielectrophoresis indicates motion towards low field zones. In our system, since the dielectric constant of the capsules is much lower than that of the surrounding medium (water), the capsules behave as dielectric `holes' and consequently the high field zone appears perpendicular to the applied field. As seen in Fig. \ref{assembly_e}j, under an applied electric field in z, and a capsule with its long axis oriented along z, the equatorial region of the capsule shows the highest field values, while the lowest values are in the poles. Consequently, at frequencies lower than 10 kHz, the capsules will tend to stay in the x-y plane (perpendicular to the field). Additionally, due to their shape, they will experience a torque. Emerse \textit{et al.} \cite{emerse2024morphologies} have recently shown that for an ellipsoid with aspect ratio $b/a$, the expression for torque generated by an applied electric field contains the frequency dependent part $g(\omega)$ where:
\begin{equation}
g(\omega)=\frac{\pi a b^2}{3}Re[K_{II}(\omega) K_{\perp}(\omega)]
\end{equation}
Depending on this factor $Re[K_{II}(\omega) K_{\perp}(\omega)]$, the torque might change its sign beyond a certain frequency. Calculation of this factor for our system is shown in Fig. \ref{assembly_e}i for $\sigma_p=10^{-4}$ S/m and $\sigma_p=10^{-6}$ S/m. For $\sigma_p=10^{-4}$ S/m we see a reversal in the sign of this factor beyond 6 kHz and one more beyond 50 kHz, but the value of the factor after the second transition is weaker compared to the first transition and saturates at higher frequencies. No such change of sign is observed for $\sigma_p=10^{-6}$ S/m. 

Our analysis indicates that in the absence of any applied field, Brownian motion makes the capsules randomly oriented on the x-y plane (low gravitational height due to high density of hematite), and no clustering/chaining is observed from the residual magnetic moment (Fig. \ref{assembly_e}a). When a field of frequency 5 kHz is applied, the capsules experience a negative torque in the z direction leading to them lying flat on the observation plane (Fig. \ref{assembly_e}b). Beyonf 5 kHz and till 50 kHz, the capsules experience a positive torque in the z direction leading to their alignment along the field direction (Fig. \ref{assembly_e}c). Between 5 kHz and 50 kHz, EHD plays a secondary role in our system, while the net effect of the electric dipolar force is to create a repulsion between either the x-y oriented capsules (at 5 kHz) or the z-oriented capsules (at 50 kHz), so that the role of the magnetic dipolar attractions is subdued. However, at 1 MHz (Fig. \ref{assembly_e}d), the electric dipolar repulsion weakens, as well as the EHD forces and as a result the capsules can come closer. The weakening of the electric dipolar force is evident from the decay in the absolute value of both $K_{II}$ and $K_{\perp}$ as seen in Fig. \ref{assembly_e}e-f. Due to this, the attractive interactions between the magnetic dipole moment vectors situated along the waist of the capsules dominate and the capsules tend to assemble into clustered bundles. Conceptually, this reorientation is reminiscent of a Fréedericksz transition in nematic liquid crystals, where a director field remains pinned in its initial alignment below a critical field strength and then reorients abruptly once that threshold is exceeded \cite{Freedericksz1933, Daftari2026}; here, an analogous threshold behavior emerges in frequency rather than field strength, and governs the orientation of individual capsules.

\begin{figure}[h]
\centering
  \includegraphics[width=8.0cm]{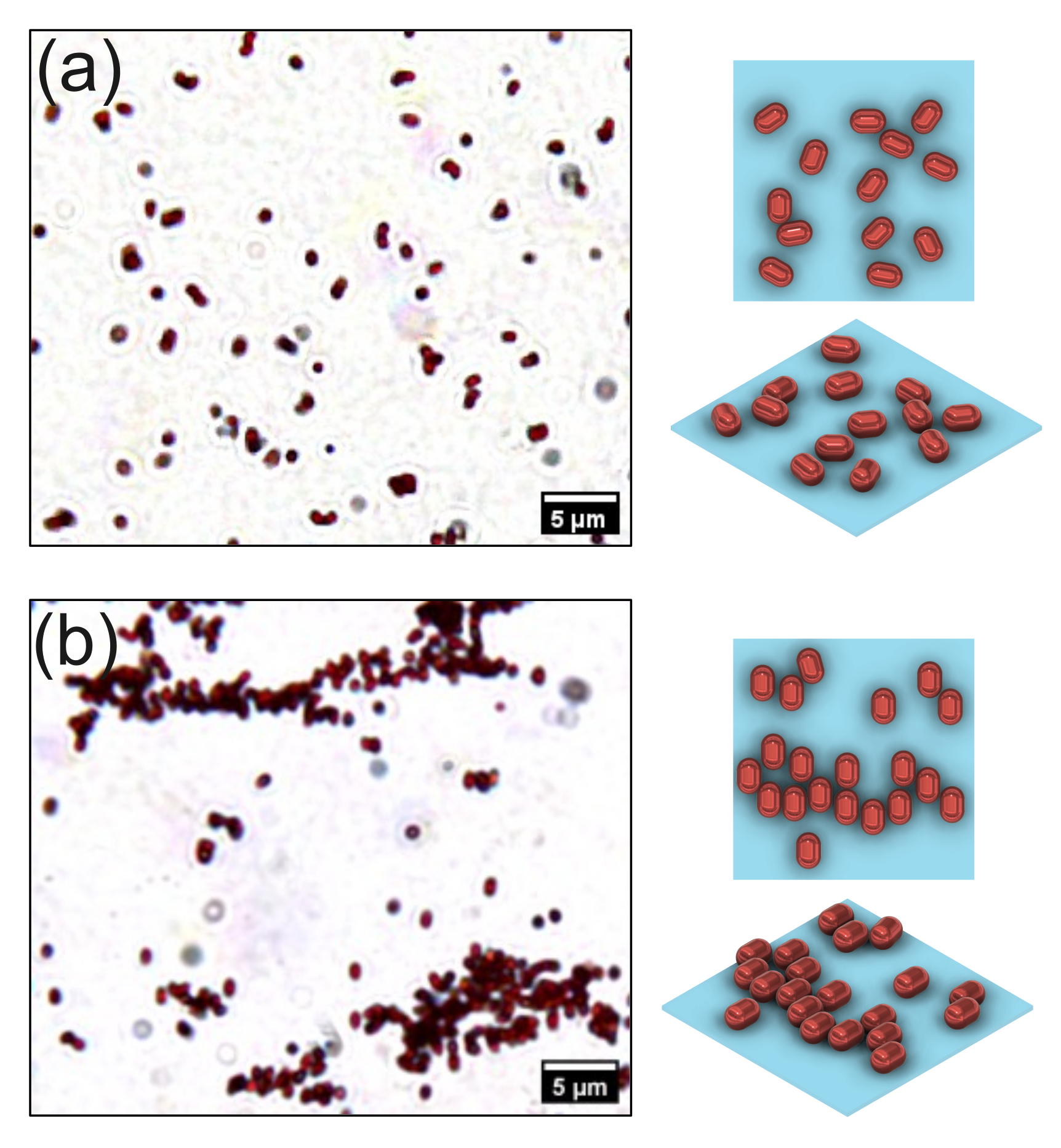}
  \caption{Assembly of the capsules under a DC magnetic field of 1.7 mT: (a) Randomly oriented, well-separated capsules when the field is applied perpendicular to the viewing plane; (b) formation of side-by-side (along the waist) ribbons, when the field is applied along the viewing plane.}
  \label{assembly_b}
\end{figure}

\begin{figure*}[!h]
\centering
  \includegraphics[width=15.0 cm]{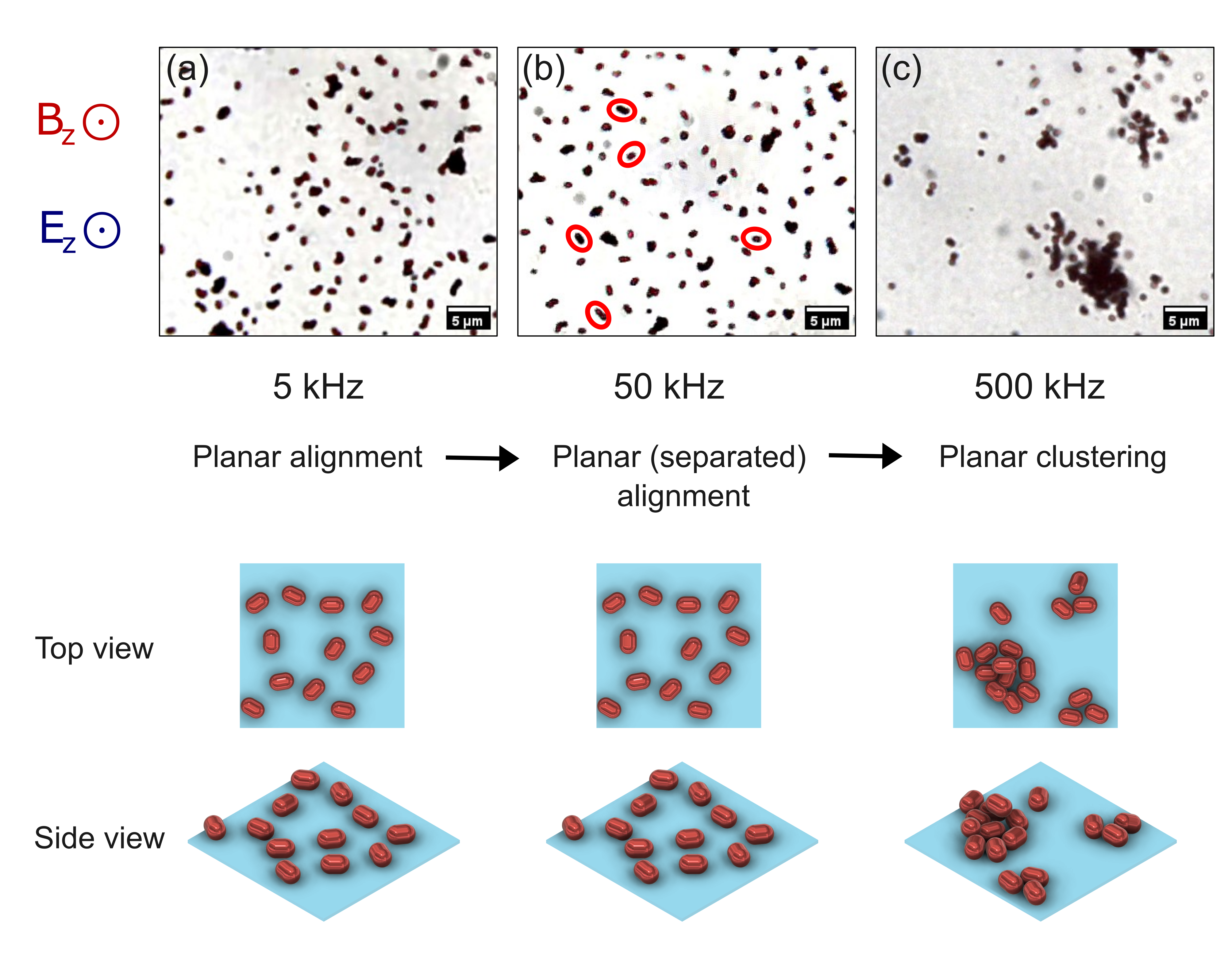}
  \caption{Assembly of the hematite capsules at $E$= 20 Vpp and $B$ = 1.7 mT with $E$ and $B$ in parallel (along z direction).(a) At 5 kHz the capsules are oriented with their long axes lying in the x-y plane. (b) At 50 kHz the planar alignment persists, but the separation between the capsules increases. (c) At 500 kHz bundling of the capsules are seen but their long axis orientation still remains along the x-y plane indicating mutually parallel alignment of the magnetic dipole moments.}
  \label{eb-field para assembly}
\end{figure*}

\subsection{Assembly of hematite capsules under Magnetic Fields}
The magnetic dipolar interactions between two magnetic point dipoles under a steady, uniform magnetic field $B$ is given as: 
\begin{equation}
    \textbf{F}_{B}(\textbf{r})=\frac{3\mu_0m^2}{4\pi r^4}\left[(1-3\cos^2{\theta})\hat{r}+\sin{2\theta}\hat{\theta}\right] 
\end{equation}
where $m=$ is the magnetic dipole moment and $\mu_0$ is the magnetic permeability of free space. For a material whose principal axes of magnetic anisotropy are along the principal axes of the ellipsoid, Osborn gave the magnetization along the ith principal axis $M_i$ as:\cite{osborn1945demagnetizing}
\begin{equation}
   M_i=\frac{\chi_i}{1+N_i\chi_i}H_{0,i} 
\end{equation}
where $M_i$ is the magnetization related to the dipole moment as $m_i=VM_i$, ans $\chi_i$ and $N_i$ are the susceptibility and demagnetization factors along the ith principal axis. Under an applied magnetic field , our hematite capsules were observed to rotate such that their short axes align parallel to the magnetic field vector. This implies that the easy axis of magnetization for the capsules is along their minor axis (waist region) similar to what has been reported previously for hematite ellipsoids of various aspect ratios.\cite{martinez-10.1039/c5sm02823j} Therefore, when a magnetic field of 1.7 mT is applied along the z direction, the in-plane magnetic dipolar repulsion causes the capsules to orient with their long axis along the x-y plane but otherwise oriented randomly (Fig. 4a). However, under a magnetic field in the x-y plane, the capsules align side by side along their minor axes, thereby forming a ribbon like structural configuration. The ribbons disintegrate rapidly on removal of the applied field, in spite of the remanent magnetization of the hematite capsules.  

\subsection{Assembly under combined Electric and Magnetic fields}
Since the hematite capsules possess both magnetic and electric dipole moments, a multi-field excitation of the capsules is expected to create a subtle interplay between the electric and magnetic forces. This can manifest as competition, cooperation or completely independent contributions of the two fields, thereby creating different assembly pathways. To examine this in detail, we considered two geometries: a) parallel and b) perpendicular application of electric and magnetic fields.

\subsubsection{Assembly under parallel electric and magnetic fields}
\begin{figure*}[h]
\centering
  \includegraphics[width=15.0 cm]{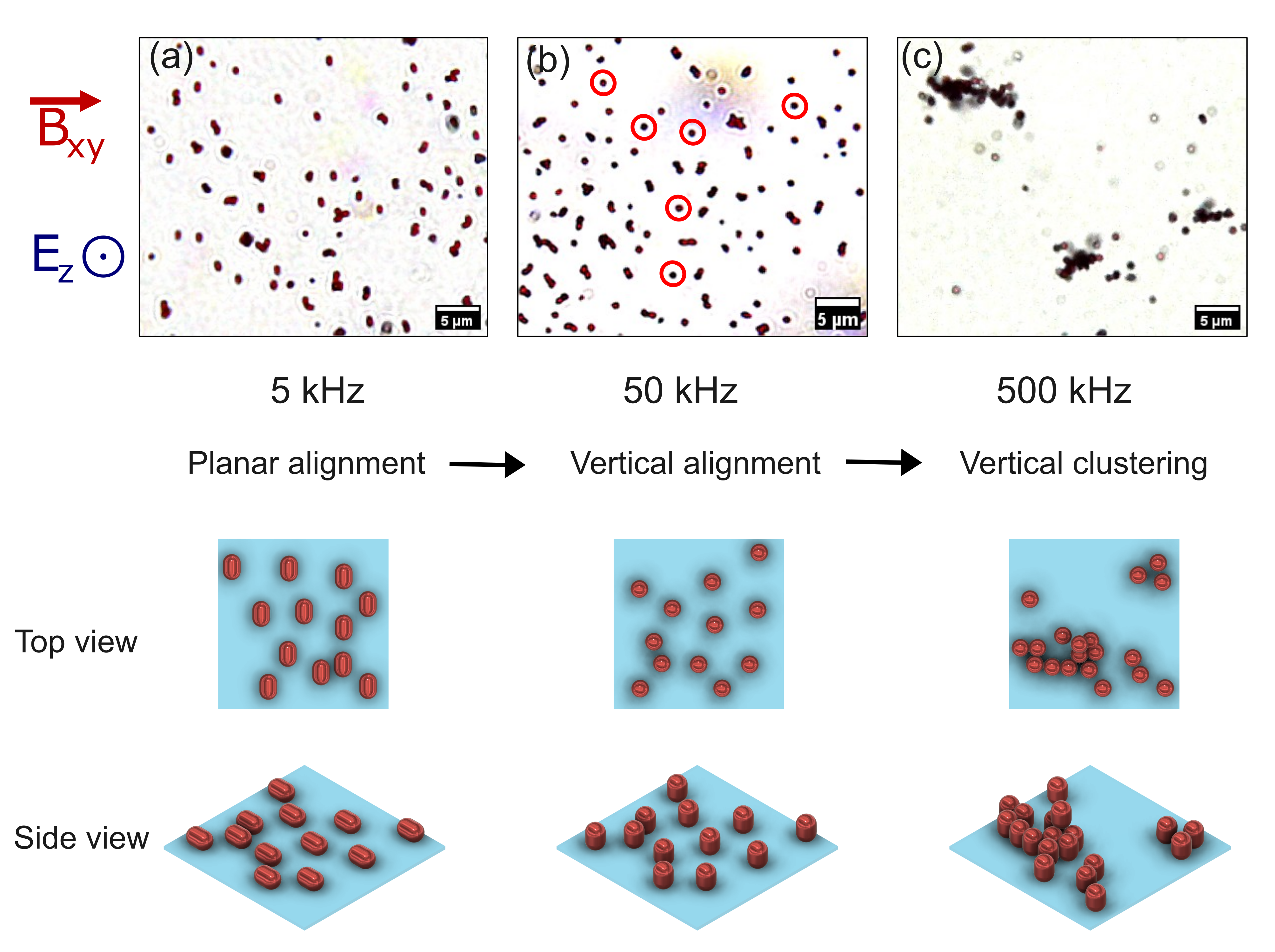}
  \caption{Assembly of the hematite capsules at $E$= 20 Vpp and $B$ = 1.7 mT with $E$ along z $B$ in the x-y plane.(a) At 5 kHz the capsules are oriented with their long axes lying in the x-y plane and their short axes on average point towards the direction of the $B$ field. (b) At 50 kHz the long axes of the capsules orient along the z axis while the magnetic dipole moment vectors along their waist are still oriented along the magnetic field direction. (c) At 500 kHz bundling of the vertically aligned capsules are seen, instead of any side-by-side ribbons.}
  \label{eb-field perp assembly}
\end{figure*}
When the electric (E) and magnetic fields (H) are applied in parallel (along the z direction), a combined effect of the electrical and magnetic torques is observed leading to the formation of new structural configurations at the same frequency/amplitude values. At 5 kHz, the capsules are oriented randomly on the x-y plane (Fig. \ref{eb-field para assembly}a) as in the case of zero B field (Fig. \ref{assembly_e}b). In this case, the electric and magnetic torques both act in cooperation to orient the long axis of the capsules parallel to the xy plane. However, at 50 kHz, while the electric torque tries to orient the capsules along the field direction (z-axis) similar to the case in Fig. \ref{assembly_e}c, the magnetic torque tends to align the magnetic dipole moment vectors of the capsules along the field direction, thereby orienting them along the x-y plane (Fig. \ref{eb-field para assembly}b). In this competition, the magnetic torque wins out over the electric torque at $B=$ 1.7 mT and $E=$20 Vpp. Additionally, the in-plane magnetic repulsion between the capsules reinforces the in-plane electric dipolar repulsion, leading to the formation of a separated planar configuration with a larger average distance between the nearest neighbors as compared to 5 kHz. Finally, at 50 kHz, surprisingly, the formation of clustered bundles similar to the case of zero $B$ field is seen, but the individual capsules are now oriented in the x-y plane, instead of being vertically oriented. Although the electric dipolar repulsion is weakened at 1 MHz, the magnetic dipolar repulsion is still the same as in 50 kHz, so the clustering of the bundles is unexpected. This might happen due to other factors like electrothermal flow at high frequencies, in which case inhomogeneous Joule heating around a particle generates convective fluid flow bringing the particles together. Tracer particle based flow studies are required to confirm this, but this is beyond the scope of this paper. To summarize, for parallel $E$ and $B$ geometry, the assembly of the capsules proceeds through a cooperation between the electric and magnetic torques at low frequencies, and a competition between them at higher frequencies. 
\subsubsection{Assembly   under perpendicular electric and magnetic fields}
When the electric field is applied along the z axis and the magnetic field is applied in the x-y plane, the capsules show a very different behavior as compared to $E$ and $B$ both along z axis. Now at 5 kHz, they orient with their long axes parallel to the x-y plane as before (Fig. \ref{eb-field perp assembly}a), but because of the in-plane magnetic field, the magnetic moment vectors along their waist all point along the field direction, so on average the long axes of the capsules have a preferred orientation on the x-y plane. Formation of side-by-side ribbons (Fig. \ref{assembly_b}b) is not seen due to the electric dipolar repulsion between the capsules. At 50 kHz, the electric torque is able to reorient the capsules with their long axes along the z direction as there is no competition with the magnetic torque that tends to keep the minor axes of the capsules aligned along the x-y plane. At 500 kHz, clustered bundles with individual vertically oriented capsules similar to the case in Fig. \ref{assembly_e}d appear. We hypothesize that any side-by-side ribbon formation in this case is hampered by the electrothermal flow induced clustering of the capsules. To summarize, under perpendicular $E$ and $B$ fields, the assembly is an independent combination of the electrical and magnetic torques, so reorientation under an electric field of higher frequency can occur while the magnetic dipole moment vectors are still aligned in the magnetic field direction.
\begin{figure*}[!h]
\centering
  \includegraphics[width=13.0cm]{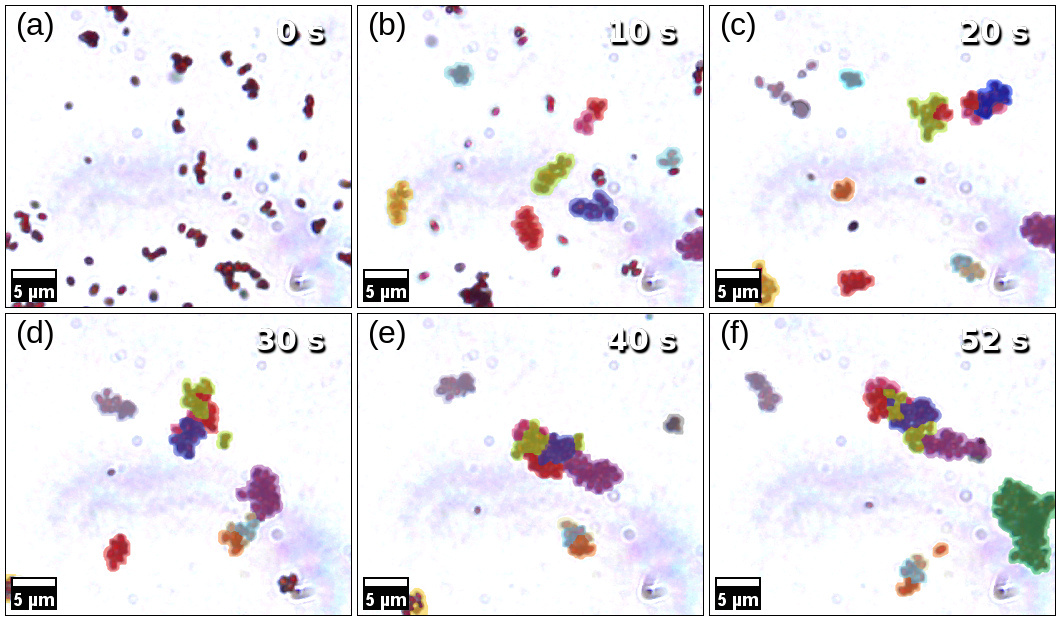}
  \caption{(a-f) Formation of dynamic, reconfigurable crystals at pH 10. The clusters are not static but rearrange, exchange particles, merge, break apart, or become unstable and explode.}
  \label{living_crystal_ph10}

  \includegraphics[width=15.0 cm]{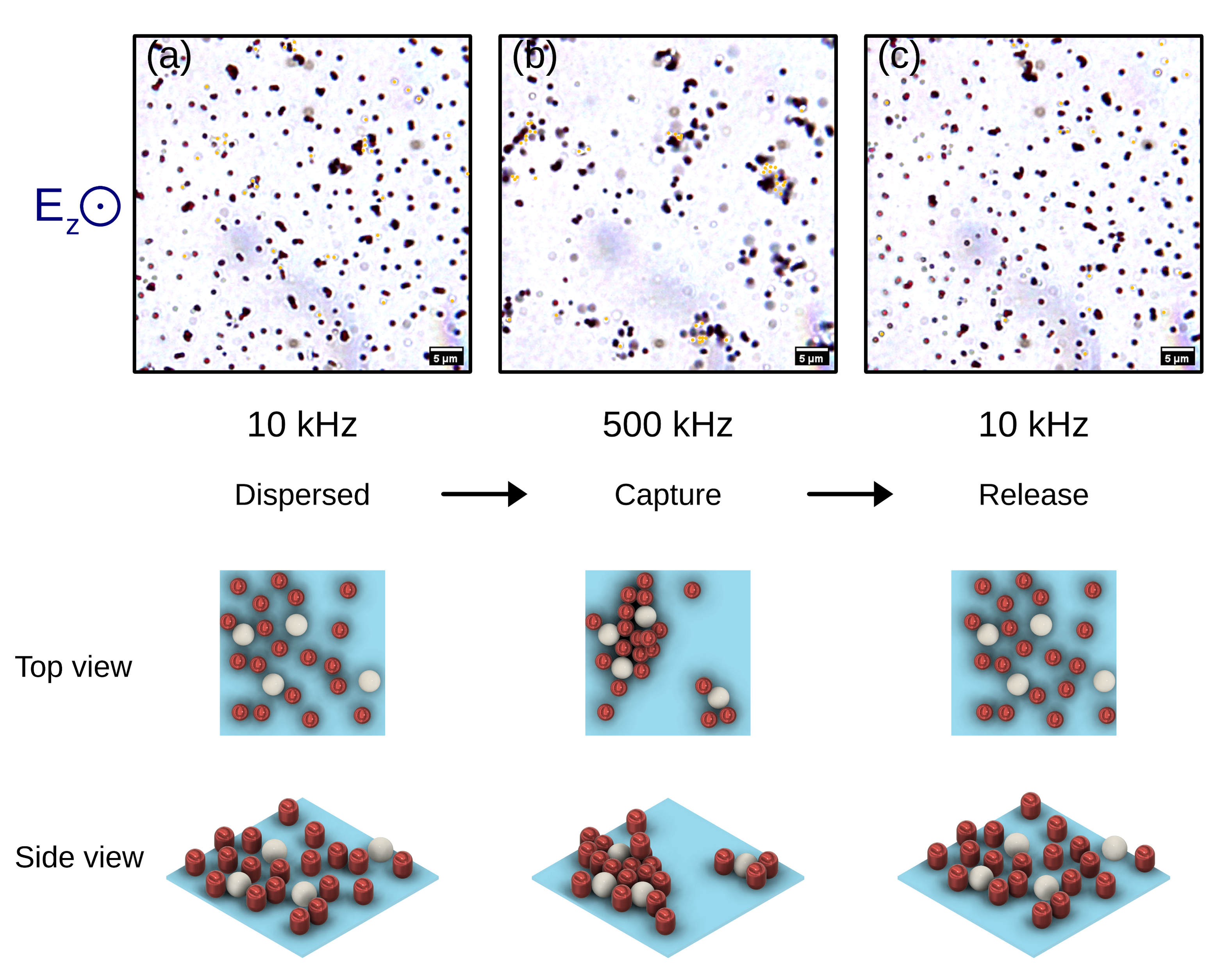}
  \caption{(a) Dispersal of Silica $(SiO_2)$ microspheres (cargo) within hematite capsules (trappers) under electric field at 50 kHz frequency, $20V_{pp}$; (b) capture of Silica $(SiO_2)$ microspheres within hematite bundles under electric field at 500 kHz frequency, $20V_{pp}$; (c) Release of the silica spheres by raising the frequency back to 50 kHz.}
  \label{sillica_trapping}
\end{figure*}

\subsection{Towards applications: Dynamic crystallization and object trapping}
In contrast to the well-separated planar alignments observed at low frequencies, the capsules undergo dynamic crystallization when the solution pH is shifted from that of Milli-Q water (between neutral and pH 5.5) to basic conditions (pH = 8-10). In a medium containing extra negatively charged ions, the electric field, and as a result the in-plane electric dipolar repulsion is screened, thereby allowing the electrohydrodynamic forces to bring the capsules into clustered configurations. The crystals formed are out-of-equilibrium dynamic configurations where particles constantly break off and join, as shown in Fig. \ref{living_crystal_ph10}a-f. The capsules at low pH resemble the 'living crystals' demonstrated by Palacci \textit{et al.}\cite{palacci2013living}, although in their case the constituent blocks were active, composite Janus particles subjected to optical excitation and ion gradients. The removal of the field will instantaneously break the crystals and revert the system to a randomly dispersed configuration. The field- and pH-induced crystallization can be used for controlled trapping and subsequent translation of cargo particles. Even without changing the pH of the MQ water, increasing the frequency to 500 kHz can cause the capsules to trap silica microspheres (shown in orange in Fig. \ref{sillica_trapping}). By changing the frequency to 50 kHz, the capsules can be instantly separated and reoriented into vertically aligned configurations, thereby releasing the silica spheres. Transport of these clusters from one point to another can be performed simply by applying a spatially inhomogeneous magnetic field. Our hematite capsules therefore offer an efficient platform with multiple control parameters for producing stimuli-responsive, reconfigurable materials, as well as microscale cargo-trapping and transport systems.

\section{Conclusions}
We studied the multi-field driven assembly of micrometer sized hematite capsules which are simultaneously dielectrics as well as canted antiferromagnets at room temperature. The morphological, crystallographic and magnetic characterization of the capsules revealed a trigonal phase, polycrystalline nature and a residual magnetization of 0.26 emu/g. Under the application of an out-of-plnae (z oriented) electric field  the capsules showed planar alignment at 5 kHz, long axes re-orientation along the z axis at 50 kHz and clustering into bundles at 500 kHz. Under a DC magnetic field in the x-y plane, the capsules clustered into side-by-side ribbons along their waist regions, indicating that the magnetic dipole moment vector is situated along their minor axes. The roles of electric and magnetic dipolar forces and torques behind these observations were discussed in detail with the help of Clausius-Mossotti factor calculations and finite element modelling. The assembly behaviour changed drastically under combined electric and magnetic fields. Under parallel E and B fields (both along z), the assembly of the capsules proceeded through a cooperation between the electric and magnetic torques at low frequencies but a competition at high frequencies. This manifested as planar alignment, separated planar alignment and planar clustering over a frequency range from 5-500 kHz. For perpendicular E and B fields (E along z, B in x-y plane) the assembly was an independent combination of the contributions of the electrical and magnetic forces/torques. This was seen from the successive planar alignment, vertical alignment and vertical clustering over the 5- 500 kHz frequency regime. By changing the pH to 10, dynamic, reconfigurable crystals could be produced even at low frequencies. The potential of the hematite capsules as cargo trappers under tunable field parameters or pH was demonstrated by capture and release of silica microparticles.

\section*{Author contributions}
JD and IB fabricated the sample chambers and conducted the electric- and magnetic-field-based experiments. JD performed the data analysis. IC conceived of the presented idea. JD, IB, and IC contributed to the planning of the experiments and writing of the paper.

\section*{Conflicts of interest}
There are no conflicts to declare.

\section*{Data availability}
Most of the data used in this work are either in the main paper or the ESI.$\dag$ Additional data will be available from the authors upon request as most of these are video data and are too bulky to store in a repository.

\section*{Acknowledgements}
We acknowledge the Start-up grant (SRG/2021/001696) of the Science and Engineering Research Board (SERB) (at present the Anusandhan National Research Foundation (ANRF)), Government of India, Additional competitive research grant (GOA/ACG/2021-2022/Nov/03) from BITS-Pilani, and MolBio research award from Molbio Diagnostics Ltd. for this work. We also acknowledge the central sophisticated instrumentation
facility (CSIF) at BITS-Goa for access to the characterization instruments.

%%%END OF MAIN TEXT%%%

%The \balance command can be used to balance the columns on the final page if desired. It should be placed anywhere within the first column of the last page.

\balance

%If notes are included in your references you can change the title from 'References' to 'Notes and references' using the following command:
%\renewcommand\refname{Notes and references}

%%%REFERENCES%%%
\bibliography{rsc} %You need to replace "rsc" on this line with the name of your .bib file

@article{Manoharan2015,
  author  = {Manoharan, Vinothan N.},
  title   = {Colloidal matter: Packing, geometry, and entropy},
  journal = {Science},
  volume  = {349},
  number  = {6251},
  pages   = {1253751},
  year    = {2015},
  doi     = {10.1126/science.1253751}
}

@article{Glotzer2007,
  author  = {Glotzer, Sharon C. and Solomon, Michael J.},
  title   = {Anisotropy of building blocks and their assembly into complex structures},
  journal = {Nature Materials},
  volume  = {6},
  pages   = {557--562},
  year    = {2007},
  doi     = {10.1038/nmat1949}
}

@article{AlHarraq2022,
  author  = {Al Harraq, A. and Choudhury, B. D. and Bharti, B.},
  title   = {Field-Induced Assembly and Propulsion of Colloids},
  journal = {Langmuir},
  volume  = {38},
  pages   = {3001--3016},
  year    = {2022},
  doi     = {10.1021/acs.langmuir.1c03378}
}

@article{Sacanna2011,
  author  = {Sacanna, Stefano and Pine, David J.},
  title   = {Shape-anisotropic colloids: building blocks for complex assemblies},
  journal = {Current Opinion in Colloid \& Interface Science},
  volume  = {16},
  pages   = {96--105},
  year    = {2011},
  doi     = {10.1016/j.cocis.2011.01.003}
}

@article{Bharti2016,
  author  = {Bharti, B. and Kogler, F. and Hall, C. K. and Klapp, S. H. L. and Velev, O. D.},
  title   = {Multidirectional colloidal assembly in concurrent electric and magnetic fields},
  journal = {Soft Matter},
  volume  = {12},
  pages   = {7747--7758},
  year    = {2016},
  doi     = {10.1039/C6SM01317A}
}

@article{Ma2013,
  author  = {Ma, F. and Wu, D. T. and Wu, Ning},
  title   = {Formation of colloidal molecules by alternating-current electric fields},
  journal = {Journal of the American Chemical Society},
  volume  = {135},
  pages   = {7839--7842},
  year    = {2013},
  doi     = {10.1021/ja402942f}
}

@article{Ma2015,
  author  = {Ma, F. and Wang, S. and Wu, D. T. and Wu, Ning},
  title   = {Electric-field-induced assembly and propulsion of chiral colloidal clusters},
  journal = {Proceedings of the National Academy of Sciences},
  volume  = {112},
  pages   = {6307--6312},
  year    = {2015},
  doi     = {10.1073/pnas.1502206112}
}

@article{Gong2017,
  author  = {Gong, Jian and Wu, Ning},
  title   = {Electric-Field Assisted Assembly of Colloidal Particles into Ordered Nonclose-Packed Arrays},
  journal = {Langmuir},
  volume  = {33},
  pages   = {5769--5776},
  year    = {2017},
  doi     = {10.1021/acs.langmuir.7b01152}
}

@article{Zhu2025,
  author  = {Zhu, X. and Wu, N.},
  title   = {The Formation and Transformation of Colloidal Clusters under Alternating-Current Electric Fields},
  journal = {Langmuir},
  volume  = {41},
  pages   = {15354--15363},
  year    = {2025}
}

@article{Barros2025SoftMatter,
  author  = {Barros, I. and Ramachandran, S. and Chakraborty, I.},
  title   = {Structure, dynamics and phase transitions in electric field assembled colloidal crystals and glasses},
  journal = {Soft Matter},
  volume  = {21},
  pages   = {1884--1894},
  year    = {2025}
}

@article{Haque2025,
  author  = {Haque, M. A. and Maestas, J. R. and Zhu, X. and Hanson, B. L. and Wu, D. T. and Wu, N.},
  title   = {High-Density and Well-Aligned Hierarchical Structures of Colloids Assembled under Orthogonal Magnetic and Electric Fields},
  journal = {ACS Nano},
  volume  = {19},
  pages   = {760--770},
  year    = {2025}
}

@article{Ketzetzi2025,
  author  = {Ketzetzi, S. and Caprini, L. and Willems, V. and Alvarez, L. and L{\"o}wen, H. and Isa, L.},
  title   = {Active Colloidal Molecules with Dynamic Configurational Freedom},
  journal = {ACS Nano},
  volume  = {19},
  pages   = {29430--29439},
  year    = {2025}
}

@article{Wei2026,
  author  = {Wei, M. and Kraft, D. J.},
  title   = {Life-like behavior emerging in active and flexible microstructures},
  journal = {Proceedings of the National Academy of Sciences},
  volume  = {123},
  pages   = {e2531743123},
  year    = {2026}
}

@article{Martchenko2016,
  author  = {Martchenko, I. and Crassous, J. J. and Mihut, A. M. and Bialik, E. and Hirt, A. M. and Rufier, C. and Menzel, A. and Dietsch, H. and Linse, P. and Schurtenberger, P.},
  title   = {Anisotropic magnetic particles in a magnetic field},
  journal = {Soft Matter},
  volume  = {12},
  pages   = {8755--8767},
  year    = {2016}
}

@article{Liao2022,
  author  = {Liao, Z. and others},
  title   = {Shape Matters in Magnetic-Field-Assisted Assembly of Prolate Colloids},
  journal = {ACS Nano},
  volume  = {16},
  pages   = {2558--2568},
  year    = {2022},
  doi     = {10.1021/acsnano.1c09208}
}

@article{Crassous2014,
  author  = {Crassous, J. J. and Mihut, A. M. and Wernersson, E. and Pfleiderer, P. and Vermant, J. and Linse, P. and Schurtenberger, P.},
  title   = {Field-induced assembly of colloidal ellipsoids into well-defined microtubules},
  journal = {Nature Communications},
  volume  = {5},
  pages   = {5516},
  year    = {2014},
  doi     = {10.1038/ncomms6516}
}

@article{Freedericksz1933,
  author  = {Freedericksz, V. and Zolina, V.},
  title   = {Forces causing the orientation of an anisotropic liquid},
  journal = {Transactions of the Faraday Society},
  year    = {1933},
  volume  = {29},
  pages   = {919--930},
  doi     = {10.1039/TF9332900919}
}

@article{Daftari2026,
  author  = {Daftari, Jehan and Barros, Indira and Chakraborty, Indrani},
  title   = {Bringing liquid crystals to the undergraduate lab: a liquid crystal-based temperature sensor and a one-pixel liquid crystal display},
  journal = {European Journal of Physics},
  year    = {2026},
  volume  = {47},
  number  = {2},
  pages   = {025502},
  doi     = {10.1088/1361-6404/ae42ff}
}

@article{barros2026field,
  title={Field Directed Assembly of Anisotropic Composite Pacman Colloids into Reconfigurable Architectures},
  author={Barros, Indira and Songara, Isha and Chakraborty, Indrani},
  journal={arXiv preprint arXiv:2610.32872},
  year={2026}
}

@article{barros2026cooperation,
  title={Cooperation, competition and multi-scale reorganisation: reconfigurable assembly of spherical colloids under combined electric and magnetic fields},
  author={Barros, Indira and Ramachandran, Sayanth and Chakraborty, Indrani},
  journal={Nanoscale},
  year={2026},
  publisher={The Royal Society of Chemistry}
}

@article{palacci2013living,
  title={Living crystals of light-activated colloidal surfers},
  author={Palacci, Jeremie and Sacanna, Stefano and Steinberg, Asher Preska and Pine, David J and Chaikin, Paul M},
  journal={Science},
  volume={339},
  number={6122},
  pages={936--940},
  year={2013},
  publisher={American Association for the Advancement of Science}
}

@book{jones1995electromechanics,
  title={Electromechanics of particles},
  author={Jones, Thomas B},
  volume={46},
  year={1995},
  publisher={Cambridge, New York}
}

@article{emerse2024morphologies,
  title={Morphologies of electric-field-driven cracks in dried dispersions of ellipsoids},
  author={Emerse, Megha and Lama, Hisay and Basavaraj, Madivala G and Singh, Rajesh and Satapathy, Dillip K},
  journal={Physical Review E},
  volume={109},
  number={2},
  pages={024604},
  year={2024},
  publisher={APS}
}

@article{ristenpart2004assembly,
  title={Assembly of colloidal aggregates by electrohydrodynamic flow: Kinetic experiments and scaling analysis},
  author={Ristenpart, WD and Aksay, Ilhan A and Saville, DA},
  journal={Physical Review E},
  volume={69},
  number={2},
  pages={021405},
  year={2004},
  publisher={APS}
}

@article{osborn1945demagnetizing,
  title={Demagnetizing factors of the general ellipsoid},
  author={Osborn, John A},
  journal={Physical review},
  volume={67},
  number={11-12},
  pages={351},
  year={1945},
  publisher={APS}
}

@article{landers2025clinically,
  title={Clinically ready magnetic microrobots for targeted therapies},
  author={Landers, Fabian C and Hertle, Lukas and Pustovalov, Vitaly and Sivakumaran, Derick and Oral, Cagatay M and Brinkmann, Oliver and Meiners, Kirstin and Theiler, Pascal and Gantenbein, Valentin and Veciana, Andrea and others},
  journal={Science},
  volume={390},
  number={6774},
  pages={710--715},
  year={2025},
  publisher={American Association for the Advancement of Science}
}

@article{shelke2025self,
  title={Self-assembly pathways towards floppy colloidal square lattices},
  author={Shelke, Yogesh and Pearce, Daniel JG and Kraft, Daniela J},
  journal={Nature Communications},
  volume={17},
  number={1},
  pages={1040},
  year={2025},
  publisher={Nature Publishing Group UK, London}
}

@article{van2024dna,
  title={DNA-Based Replication of Programmable Colloidal Assemblies},
  author={van Kesteren, Steven and Diethelm, Pascal and Jung, Se-Hyeong and Isa, Lucio},
  journal={Small},
  volume={20},
  number={35},
  pages={2400180},
  year={2024},
  publisher={Wiley Online Library}
}

@article{rogers2016using,
  title={Using DNA to program the self-assembly of colloidal nanoparticles and microparticles},
  author={Rogers, W Benjamin and Shih, William M and Manoharan, Vinothan N},
  journal={Nature Reviews Materials},
  volume={1},
  number={3},
  pages={16008},
  year={2016},
  publisher={Nature Publishing Group}
}

@article{chakraborty2022self,
  title={Self-assembly dynamics of reconfigurable colloidal molecules},
  author={Chakraborty, Indrani and Pearce, Daniel JG and Verweij, Ruben W and Matysik, Sabine C and Giomi, Luca and Kraft, Daniela J},
  journal={ACS nano},
  volume={16},
  number={2},
  pages={2471},
  year={2022}
}

@article{rinaldin2019colloid,
  title={Colloid supported lipid bilayers for self-assembly},
  author={Rinaldin, Melissa and Verweij, Ruben W and Chakraborty, Indrani and Kraft, Daniela J},
  journal={Soft Matter},
  volume={15},
  number={6},
  pages={1345--1360},
  year={2019},
  publisher={The Royal Society of Chemistry}
}

@article{aubret2018targeted,
  title={Targeted assembly and synchronization of self-spinning microgears},
  author={Aubret, Antoine and Youssef, Mena and Sacanna, Stefano and Palacci, J{\'e}r{\'e}mie},
  journal={Nature Physics},
  volume={14},
  number={11},
  pages={1114--1118},
  year={2018},
  publisher={Nature Publishing Group UK, London}
}

@article{zhao2026effect,
  title={Effect of a Cone Shape on the Motion of Active Janus Colloids},
  author={Zhao, Zhiyuan and Verouden, Teun WJ and Mohapatra, Dillip K and Simons, Evert ME and Meijer, Janne-Mieke},
  journal={arXiv preprint arXiv:2606.16391},
  year={2026}
}

@article{kwaadgras2014orientation,
  title={Orientation of a dielectric rod near a planar electrode},
  author={Kwaadgras, Bas W and Besseling, Thijs H and Coopmans, Tim J and Kuijk, Anke and Imhof, Arnout and Blaaderen, Alfons van and Dijkstra, Marjolein and Roij, Ren{\'e} van},
  journal={Physical Chemistry Chemical Physics},
  volume={16},
  number={41},
  pages={22575--22582},
  year={2014},
  publisher={The Royal Society of Chemistry}
}

@article{edwards2014controlling,
  title={Controlling colloidal particles with electric fields},
  author={Edwards, Tara D and Bevan, Michael A},
  journal={Langmuir},
  volume={30},
  number={36},
  pages={10793--10803},
  year={2014},
  publisher={ACS Publications}
}

@article{yang2019impact,
  title={The impact of stern-layer conductivity on the electrohydrodynamic flow around colloidal motors under an alternating current electric field},
  author={Yang, Xingfu and Johnson, Sophia and Wu, Ning},
  journal={Advanced Intelligent Systems},
  volume={1},
  number={8},
  pages={1900096},
  year={2019},
  publisher={Wiley Online Library}
}

@article{kamal2020path,
  title={Path-dependent self-assembly of magnetic anisotropic colloidal peanuts},
  author={Kamal, Md Arif and Petukhov, Andrei V and Pal, Antara},
  journal={The Journal of Physical chemistry. B},
  volume={124},
  number={27},
  pages={5754},
  year={2020}
}

@article{rossi2018self,
  title={Self-organization in dipolar cube fluids constrained by competing anisotropies},
  author={Rossi, Laura and Donaldson, Joe G and Meijer, Janne-Mieke and Petukhov, Andrei V and Kleckner, Dustin and Kantorovich, Sofia S and Irvine, William TM and Philipse, Albert P and Sacanna, Stefano},
  journal={Soft Matter},
  volume={14},
  number={7},
  pages={1080--1087},
  year={2018},
  publisher={The Royal Society of Chemistry}
}

@article{sudhakaran2020orientation,
  title={Orientation dependent interaction and self-assembly of cubic magnetic colloids in a nematic liquid crystal},
  author={Sudhakaran, Devika Venkuzhy and Pujala, Ravi Kumar and Dhara, Surajit},
  journal={Advanced Optical Materials},
  volume={8},
  number={7},
  pages={1901585},
  year={2020},
  publisher={Wiley Online Library}
}

@article{sacanna2013engineering,
  title={Engineering shape: the novel geometries of colloidal self-assembly},
  author={Sacanna, Stefano and Pine, David J and Yi, Gi-Ra},
  journal={Soft matter},
  volume={9},
  number={34},
  pages={8096--8106},
  year={2013},
  publisher={The Royal Society of Chemistry}
}

@article{sacanna2012magnetic,
  title={Magnetic click colloidal assembly},
  author={Sacanna, Stefano and Rossi, Laura and Pine, David J},
  journal={Journal of the American Chemical Society},
  volume={134},
  number={14},
  pages={6112--6115},
  year={2012},
  publisher={ACS Publications}
}

@article{tierno2014recent,
  title={Recent advances in anisotropic magnetic colloids: realization, assembly and applications},
  author={Tierno, Pietro},
  journal={Physical chemistry chemical physics},
  volume={16},
  number={43},
  pages={23515--23528},
  year={2014},
  publisher={The Royal Society of Chemistry}
}

@article{sugimoto1998shape,
  title={Shape control of monodisperse hematite particles by organic additives in the gel--sol system},
  author={Sugimoto, Tadao and Itoh, Hiroyuki and Mochida, Takeaki},
  journal={Journal of Colloid and Interface Science},
  volume={205},
  number={1},
  pages={42--52},
  year={1998},
  publisher={Elsevier}
}

@article{cantedPhysRev.116.1447,
  title = {Magnetic Properties of Hematite Single Crystals. I. Magnetization Isotherms, Antiferromagnetic Susceptibility, and Weak Ferromagnetism of a Natural Crystal},
  author = {Lin, S. T.},
  journal = {Phys. Rev.},
  volume = {116},
  issue = {6},
  pages = {1447--1452},
  numpages = {0},
  year = {1959},
  month = {Dec},
  publisher = {American Physical Society},
  doi = {10.1103/PhysRev.116.1447},
  url = {https://link.aps.org/doi/10.1103/PhysRev.116.1447}
}

@article{martinez-10.1039/c5sm02823j,
    author = {Martinez-Pedrero, Fernando and Cebers, Andrejs and Tierno, Pietro},
    title = {Orientational dynamics of colloidal ribbons self-assembled from microscopic magnetic ellipsoids},
    journal = {Soft Matter},
    volume = {12},
    number = {16},
    pages = {3688-3695},
    year = {2016},
    month = {04},
    issn = {1744-683X},
    doi = {10.1039/c5sm02823j},
    url = {https://doi.org/10.1039/c5sm02823j},
    eprint = {https://pubs.rsc.org/sm/article-pdf/12/16/3688/4616379/c5sm02823j.pdf}
    }
\bibliographystyle{rsc} %the RSC's .bst file

\end{document}